\documentclass[conference]{IEEEtran}
\IEEEoverridecommandlockouts
\usepackage{cite}
\usepackage{amsmath,amssymb,amsfonts}
\usepackage{algorithmic}
\usepackage{graphicx}
\usepackage{textcomp}
\usepackage{xcolor}
\def\BibTeX{{\rm B\kern-.05em{\sc i\kern-.025em b}\kern-.08em
    T\kern-.1667em\lower.7ex\hbox{E}\kern-.125emX}}
\begin{document}

\title{The Everyday Security of Living with Conflict}

\author{\IEEEauthorblockN{Jessica McClearn}
\IEEEauthorblockA{\textit{Royal Holloway, University of London}}
\IEEEauthorblockN{jessica.mcclearn.2021@live.rhul.ac.uk}\\
\and
\IEEEauthorblockN{Reem Talhouk}
\IEEEauthorblockA{\textit{Northumbria University}}
\IEEEauthorblockN{reem.talhouk@northumbria.ac.uk}\\
\and
\IEEEauthorblockN{Rikke Bjerg Jensen}
\IEEEauthorblockA{\textit{Royal Holloway, University of London}}
\IEEEauthorblockN{rikke.jensen@rhul.ac.uk}

}

\maketitle

\begin{abstract}
When `cyber' is used as a prefix, attention is typically drawn to the technological and spectacular aspects of war and conflict -- and, by extension, security. We offer a different approach to engaging with and understanding security in such contexts, by foregrounding the everyday -- mundane -- experiences of security within communities living with and fleeing from war. We do so through three vignettes from our field research in Colombia, Lebanon and Sweden, respectively, and by highlighting the significance of ethnography for security research with communities living in regions afflicted by war. We conclude by setting out a call to action for security researchers and practitioners to consider such lived experiences in the design of security technology that aims to cater to the needs of communities in `global conflict and disaster regions'.
\end{abstract}


\maketitle

\section{MUNDANITY IN THE SPECTACULAR}

\noindent Images of the destruction and suffering caused by wars waged, often at a distance, make their way to our screens through different news platforms. We witness smoke billowing from dropped bombs and people searching for missing family members under the rubble, alongside stories of the scattering of people as they are displaced. We are presented with reports on how advanced technologies enable `targeted operations' and `precision bombing'; yet, little reflection is offered on how such mechanisms of (modern) warfare impact communities that experience conflict at close proximity. This is not surprising, however, with much security-driven research concentrating on technological advances and policy interventions (e.g.~international law) in the context of conflict. With the foregrounding of \emph{spectacular} and \emph{technological} depictions of war, security considerations are often reduced to questions related to its \emph{cyber} capabilities. Typically, this means that the security concerns for those living in conflict and disaster regions are overlooked; their security is generally not reliant on the security of the technology itself, i.e.~\emph{cyber} security, but on the security that the technology may afford them to establish a sense of continuity and control in an everyday shattered by conflict and fears. In other words, while technology might serve a security purpose for communities living through conflict, this purpose often lies in its social and relational faculties more than the security of the technology itself. It is in this mundane that we can also observe what security looks and feels like for those impacted by war. To do so, we argue, requires a broad conception of security; a conception that brings computer security into conversation with socially grounded and immersive research approaches, such as ethnography, enabling insights and interpretations of how security and technology are interwoven. 

Building on our previous field research we show how, in contexts of conflict, it is often the \emph{mundane} daily routines and practices that shape people's understanding of security and how technology shapes it. We thus call for research and practice that broadens understandings of security in conflict settings by grounding security in how people go about their daily lives. We ask: \emph{How do people living through conflict navigate daily experiences of insecurity, and what are the security needs of those displaced by conflict?} Many of the answers to such questions can be found by observing how people practice security in their everyday. In the words of Crawford and Hutchinson: 

\begin{quote}
\dots [T]he `everyday' acts as an important counter to a prevailing emphasis upon the `spectacular' and the `exceptional', which cast a long shadow over security research~\cite[p.5]{BJC:CraHut16}.
\end{quote}

\noindent Drawing upon \emph{everyday security}~\cite{BJC:CraHut16,IPS:Nyman21} we demonstrate, through three vignettes from our field research in Colombia (2024), Lebanon (2022) and Sweden (2018), how security for people in specific conflict contexts is rooted in their everyday social and relational ties. We therefore also insist on understanding the security concerns of people living through conflict by engaging with the mundane, rather than through sensationalist narratives or by foregrounding technological advances as are often captured in the prefix \emph{cyber}. This also means that throughout this article we consider what security means -- and what it is -- to the participants in our vignettes. Thus, we do not approach security from the understanding that it can be neatly separated into \emph{either} `social' \emph{or} `technological', but argue for a conception where it can be -- and often is -- both. Our project is founded in a call to action; a call for security researchers, practitioners and designers to step into the social realities and lived security experiences of those living through, and displaced by, conflict. As noted by the authors of~\cite{FCJ:AGRS15}, this is in contrast to the prevailing practice within computer security, where ``in the absence of far away users under threat, designers can invoke them at will and imagine their needs''. To do so requires a methodological approach that enables immersion in the contexts and communities under study: ethnography. 

\section{EVERYDAY SECURITY IN CONFLICT}

\noindent The very nature of war is of course catastrophic; yet, the sensationalising of war disguises the lived reality and day-to-day experiences of communities living through it. Similarly, when \emph{cyber security} is invoked in the context of war, attention is typically given to narratives of \emph{cyber policy}, \emph{cyber warfare} and nation state \emph{cyber capabilities}. Similarly, Hutchings~\cite{SP:Huthings24} argues how the language of war has captured narratives of \emph{cyber crime}, which has led to sensationalist depictions that do not reflect the ``lived reality'' of those impacted by it. Further, the authors of~\cite{vu2024getting} in their work on the nature and frequency of recent DDoS attacks between Russia and Ukraine, highlight how the spectre of \emph{cyber war} masks the actual technical reality of such attacks. Such narratives risk obfuscating the more mundane experiences of communities living through war, where security is often rooted in daily routine activities and practices that provide a sense of stability and predictability -- and, by extension, security. 

We position our contribution within our own prior co-authored work -- and the work of others -- that recognises the significance of the everyday in security research. For example, in~\cite{CHI:ColJen19,CHI:ColJenTal18,CSCW:JenColTal20} we used the framing of `the everyday' to analyse how social aspects shaped digital security for refugees resettling in a new land. In~\cite{CSCW:TCJBGGAAM20} we highlighted the everyday insecurities experienced by Syrian refugees in Lebanon navigating food aid technologies, while we in~\cite{mcclearn2023othered, mcclearn2024security} showed the everyday experiences of insecurity in Lebanon among marginalised groups and how they tried to overcome infrastructural failures. Individually -- and collectively -- these works highlight how the technological needs of communities living through or displaced by conflict are rooted in mundane practices of security, such as the sourcing of food, navigating broken infrastructures, learning how to take a bus, comprehending new cultural contexts, and searching for a job and housing. Therefore, although the participants in these studies relied on technology to navigate their everyday security concerns, these concerns were not technological. 

Using our individual voices and experiences of conducting security research in different contexts, we provide vignettes from our fieldwork in Lebanon, Colombia and Sweden. We do so to illustrate how (in)security was woven into the fabric of everyday life (and survival) for the people we encountered in these three settings, showing how and for what purpose the people we encountered relied on technology for their security and, in particular, the security of others. Thus, staying with our call to action, our vignettes suggest that designing for those living through or displaced by conflict requires understanding not simply their use of technology but the mundane contexts within which they use technology and the reasons they give.

\section{Vignette: Empty houses, Lebanon}

\noindent The ongoing conflict in Lebanon has led to an economic crisis which is evidenced in the movement of populations, the breaking down of physical infrastructures and everyday security practices. This economic crisis has been referred to by the World Bank as a `ponzi scheme'~\cite{worldbank2022}, forcing many people to leave the country temporarily and seek economic security elsewhere. 

One participant,\footnote{We anonymise all participants to ensure their protection.} who had spent some time outside of Lebanon and since returned to the country, emphasised that Lebanese people \emph{``are not static''.} The participant noted how the state of flux was also visible in the built environment: \emph{``If you go to certain areas of Lebanon you will just find empty houses, empty villas that people build with the money they made abroad to show that they belong. Then for example people pass away or they never come back and the property is just empty.''} The traces of people having moved, such as empty houses, created a sense of having been left behind, of being stuck, while pointing to the displacement of many Lebanese people. 

Those who had decided to remain in Lebanon were living with the daily consequences of the conflict. For example, because of the economic instability of the country, many people lacked access to their personal finances as restrictions had been placed on ATM cash withdrawals. This meant that people in Lebanon would often rely on money transfers from the Lebanese diaspora. This paired with limited stable internet access led to a daily struggle to secure personal finances and, ultimately, to survive. These challenges of access overshadowed any concerns related to technological security or, more significantly, privacy in the context of sharing sensitive and personal details with those whose help could ensure one's survival -- through money transfers from abroad. One participant shared: 

\begin{quote}
\textit{Relatives are sending money which is the lifeline right now [\dots] We are at a point where talking about [digital] rights becomes a conversation which a lot of the public is out of the loop of, if they are worried about their immediate survival.}
\end{quote}

\noindent The transient nature of life in Lebanon had deep-rooted implications for people's everyday security; deserted buildings served as reminders of the uncertainty that continued to shape life in Lebanon. The impact of movement and temporality in the context of conflict -- being displaced for shorter or longer periods of time -- is an area of burgeoning security research which deserves further attention. How (in)security presents itself for transient communities and their reliance on specific technological functionalities, such as the wiring of money from afar, requires an understanding of the entanglement of identity and security in this setting. Combined with the `emptiness' caused by the conflict, the struggle to survive, as shown in this vignette, left those living in Lebanon at the mercy of \emph{both} the functionalities enabled by money transfer technologies \emph{and} the Lebanese diaspora whose empty houses remained as physical reminders of their displacement. Here, security concerns, while not technological, were partly mitigated by practices enabled \emph{through} technology.

\section{Vignette: Being heard, Colombia}

\noindent Gender, land and security are intimately interwoven in the department of Cauca in Colombia. Many groups in Colombia live in rural zones, heavily controlled by paramilitary organisations. The deep-rooted gendered expectations of women to look after the home, and the scarcity of technology in such areas, left many women feeling abandoned. This was exemplified by one participant, from rural Cauca, who worked for an international non-governmental organisation (NGO) as she discussed the gendered nature of the already limited technological infrastructures in the rural zones, and how this impacted the lives of women: 

\begin{quote}
\textit{In the end, who is going to have it [a mobile phone]? The man who goes out more, and the woman who is often relegated to those private roles in the home, she does not have access to technology.}
\end{quote} 

\noindent Limited technological infrastructures paired with gender divides meant that women were often confined to the home. Labour related to the home was also extended to the surrounding land and natural resources. Illegal mining and extractive practices had contaminated many rivers, leading to health concerns which unequally impacted women. One participant stressed the significance of this: \emph{``The river is the only source for their [the women] life, for food, for fishing, for washing clothes.''} As a form of resistance, many women often became social leaders and land defenders. While this mobilisation of women meant that they had become more visible in their communities, it also led to an increase in the number of enforced disappearances of women activists. As a result, many women decided to stay silent. Another participant, a conflict victim and social leader, explained this during the fieldwork: 

\begin{quote}
\textit{Many have decided to remain silent and keep quiet, that is why our body gets sick and screams, what the mouth is silent about and what the body feels. So then we get sick from many things and it is because of the lack of listening.}
\end{quote}

\noindent For these women, the tensions between everyday security, on the one hand, and their increased visibility, on the other, were intimately tied to the continued conflict in Cauca. To translate the significance of such mundane normalities for security research and practice, this vignette demonstrates how, for these women, security was observed through their experienced lack of technological infrastructures and the prevalence of gendered risks as well as discriminatory practices. In many ways, the absence of technology in their daily struggle for survival was seen as a security risk that left many women isolated in rural settings. The vignettes thus demonstrates how there is a concrete need for security technologies that both permit and enable communities in at-risk contexts to be heard, while taking into consideration, factors such as gender, language, limited infrastructures and economic deprivation. While tensions between security and visibility in at-risk contexts are not new, here such tensions materialise through everyday forms of resistance.  

\section{Vignette: On a bus, Sweden}

\begin{quote}
\textit{Okay, so maybe sometimes the phone is not good, but right now, for me, the phone is my right hand.}
\end{quote}

\noindent This is a quote from a Syrian refugee in Sweden during fieldwork in 2018. She had been forced to leave behind her parents and husband in conflict-ridden Syria two years earlier. When she arrived in Sweden she had been granted the `right to remain'. Yet, like so many others in her situation, she was still waiting to receive a decision on her application for asylum in Sweden. Although not unique to her, her story is one of uncertainty and of being suspended between the life (and the people) she had left behind in Syria and the life she was trying to establish in Sweden -- still with the hope that her family would be able to join her \emph{soon}. 

The quote relates to her experience of sitting on a bus in Malm\"{o}, Sweden, as she was going from her temporary accommodation to her daily language classes. She never left her house without her mobile phone, ensuring that it was fully charged but still carrying the charger -- in case she would need it throughout the day, which she usually did. While displaced from her family, this daily routine gave her a sense of security; knowing that she would always be reachable should anyone in Syria need her. 

Sitting on this bus, on this day, she had received a call from her husband: \emph{``My husband and my parents were on their way to the Swedish visa application centre in Beirut from Syria, when they needed my `right to remain' in Sweden in order to be able to enter Lebanon as they were told they would not be allowed to enter without it.''} This is a document that she would always carry with her. Still on the bus, she took a picture of it and quickly shared with her husband; enabling her family to enter Lebanon and take the next steps in their own asylum-seeking processes. For her, everyday security was captured in that moment. While physically displaced by the conflict, she was closely connected to it, both through the experiences of others and her own memories from a life of navigating an everyday existence during political and social upheaval. Her mobile phone had become an extension of herself (\emph{``my right hand''}), through which she lived the security (and insecurity) of others. Thus, computer security mechanisms such as access control, privacy settings and encrypted messaging played no role in her everyday security. Security for her was to be able to quickly share and receive information -- not whether such information was protected. 

\section{A CALL TO ACTION}

\noindent It is in the \emph{mundane} that we can observe and question the tensions between security and insecurity that shape people's lived experiences in times of conflict. As our vignettes exemplify, however, \emph{mundanity} is neither monolithic nor static; it is contextual and dynamic. We therefore caution against drawing distinct conclusions from our vignettes. Rather, our aim is that these vignettes shine a light on the often intricate, subtle and diverse ways in which security is experienced -- also during conflict and war. Our call to action is threefold,

For \emph{researchers} to understand the security practices and needs of people living with and through conflict necessitates adopting methodologies that are designed to engage with \emph{the everyday} and the subtleties that a sensationalist perspective -- often by invoking a \emph{cyber} framing -- fails to capture. This also requires a broad conception of security, where the (academic) distinction between safety and security is less significant, as exemplified through our vignettes. For the participants in these vignettes, the separation between social and technological foundations of security would be equally artificial. We follow in the footsteps of others (e.g.~\cite{USENIX:SCBAPB22,USENIX:ABJM21}) who have called for a diversification of methodologies within security-driven research. Specifically, we call for immersion in `the everyday' of those living in global conflict and disaster regions, to understand what security means to them and how they reason about it -- \emph{before} considering the potential role of (security) technology in such settings. 
    
For \emph{developers} to consider what designing for displaced populations actually means there is a pressing need to engage with the communities being designed for. Further, for (security) technologies to serve communities impacted by conflict, the design of such technologies must be grounded in the mundane experiences and social reality of such communities. Thus, through practices of co-design and participatory methods, we call on developers to collaborate with community-facing security researchers who use diverse methodologies, including ethnographic methods, to study the needs of distinct communities \emph{in situ}. This, we posit, has the possibility of affording technologists a deeper understanding of the lived, social realities and material conditions of communities displaced by war, rather than imagining their needs at will -- acknowledging, as our vignettes show, how contexts and conditions are dynamic.

For \emph{practitioners} such as policy designers to engage with the mundane, we call for computer security to be situated in broader conceptions of security. Despite increasing traction around human-centered security research, security, when considered in computer security venues, is typically understood as a technological question. We argue that this focus overlooks the lived, mundane experiences of conflict and displacement. We thus challenge (as others have done) the use of the prefix: \emph{cyber} -- as it, often unhelpfully, focuses attention on the distinctly \emph{non-human} aspects of war and security. It further casts a shadow over and distorts the lived experiences of people during conflict, where security is embedded in how people go about their daily lives -- in the \emph{mundane}.
    
\section{ACKNOWLEDGMENTS}
\noindent This work would not exist without the many contributions from people, both in and outside of Lebanon, Colombia and Sweden, who generously gave their time to speak to us and trusted us with their knowledge. The research of Jessica McClearn was supported by the EPSRC as part of the Centre for Doctoral Training in Cyber Security for the Everyday at Royal Holloway, University of London (EP/S021817/1).

\bibliographystyle{IEEEtran}
\bibliography{local}

\section{Authors}
\begin{itemize}
    \item \textbf{{Jessica McClearn}}{\,} is a PhD candidate at Royal Holloway, University of London in the Centre for Doctoral Training in Cyber Security for the Everyday. She uses ethnographic methods to explore the intersection of digital and ontological security within political and interpersonal conflicts. Contact her at jessica.mcclearn.2021@live.rhul.ac.uk
    \item \textbf{{Reem Talhouk}}{\,} is an associate professor in Design and Global Development at Northumbria University. She has conducted research with refugees, activists and humanitarians, with a focus on the interplay between sociotechnical systems and mechanisms of oppression and resistance. Contact her at reem.talhouk@northumbria.ac.uk.
    \item \textbf{{Rikke Bjerg Jensen}} {\,} is a professor in the Information Security Group at Royal Holloway, University of London. She is a social scientists, an ethnographer, and focuses on the information security experiences, needs, practices and perspectives of groups of people in higher-risk and/or marginalised settings. Contact her at rikke.jensen@rhul.ac.uk.
\end{itemize}

\end{document}